\documentclass[prl,twocolumn,groupedaddress,showpacs]{revtex4}
\usepackage{graphics}
\usepackage{enumerate}
\begin{document}
\newcommand{\beq}{\begin{equation}}
\newcommand{\eeq}{\end{equation}}
\newcommand{\beqa}{\begin{eqnarray}}
\newcommand{\eeqa}{\end{eqnarray}}
\bibliographystyle{prl}

\title{OPERA, SN1987a and energy dependence of superluminal neutrino velocity}

\author{N.D. Hari Dass}
\email{dass@cmi.ac.in }
\affiliation{Chennai Mathematical Institute, Chennai \& CQIQC, Indian Institute of Science, Bangalore,
India.}


\begin{abstract} This is a brief note discussing the energy dependence of superluminal neutrino
velocities recently claimed by OPERA \cite{actone,brunetti}. The analysis is based on the data provided there
on this issue, as well as on consistency with neutrino data from SN1987a as recorded by the Kamioka
detector \cite{kamioka}. It is seen that it is quite
difficult to reconcile OPERA with SN1987a. The so called Coleman-Glashow dispersion relations
do not do that well, if applied at all neutrino energies. The so called \emph{quantum gravity inspired}
dispersion relations perform far worse. Near OPERA energies both a energy-independent velocity, as well as a linear energy dependence with an offset
that is comparable in value to the observed ${\delta v}$ by OPERA at 28.1 GeV work very well. Our analysis
shows that precision arrival time data from SN1987a still allow for superluminal behaviour for supernova
neutrinos. A smooth interpolation is given that reconciles OPERA and SN1987a quite well. It suggests a fourth
power energy dependence for $\frac{{\delta v}}{c}$ of supernova neutrinos. This behaviour is insensitive to whether
the velocities are energy-independent, or linearly dependent on energy, near OPERA scale of energies.
Suggestions are made for experimental checks for these relations.

\end{abstract}

\maketitle
\section{Introduction}
The recent announcement by the OPERA collaboration \cite{actone,brunetti} that they have measured superluminal
neutrino velocities has stunned both the scientific community, as well as public at large. The announced
magnitudes
of the effects are large and statistically very significant. The full implications, should the experiment
be vindicated in future, are mind-boggling. Everything in modern physics, as we understand it today,
depends one way or the other on the correctness of the Special Theory of Relativity. 
Interestingly, speeds greater than $c$ for photons in the same energy range as OPERA neutrinos, had been reported by Gharibyan \cite{gharibyan} before.

Because of the enormous stakes involved, every effort, both experimental and theoretical, must be made
to make sure that the claims are correct. On the experimentalists side, one needs to thoroughly check and 
double check all sources of errors. One needs to repeat the experiment, both the original version, as
well as variants of it like at different energies, different path lengths etc.. On the theoretical side
efforts have to be made to check the consistency of the OPERA data both conceptually, as well as 
with earlier precision data. Indeed, after the OPERA announcement, a number of interesting ideas have
been put forward(see \cite{theo1}-\cite{theo28}). Some of these ideas even predated the OPERA results \cite{operasn10,operasn11,theo4,theo5,theo6}.

But if the OPERA results are really right, all of physics has to be radically revised and there may not exist,
at present, any sensible way of analysing it. While one may come up with some parametrizations,
or some perturbative effective treatments like the one offered by Kostelecky and coworkers \cite{kostelecky1,kostelecky2,
diazreview,russellreview} or by
 Coleman and Glashow \cite{colemanglashow},
the inner selfconsistency of such approaches is never guaranteed till another succesful theory replaces
Special Theory of Relativity. However, theories based on \emph{Spontaneous Breaking of Lorentz Invariance} \cite{ssb1,ssb2}
may be better in this regard. That will certainly entail some radical revisions of our notions of space
and time, to say the least. That way we can liken the situation to the days of the Bohr atom, where one
could parametrize some small corrections to observables in the large quantum number states, without having
the slightest idea of the deep conceptual changes that the full quantum theory eventually brought about. So here
too one may have to develop intermediate \emph{crutches} like the correspondence principle etc to make progress.

To illustrate this point, consider a number of very interesting papers have appeared which have shown that
the OPERA result in conjunction with reasonable assumptions like energy-momentum conservation, 
can lead to contradictory results \cite{kinematics1,kinematics2,kinematics3,kinematics4,kinematics5}.
On the one hand it is shown that the kinematics of pion decay is constrained to the extent of not permitting 
neutrinos of the energy 'seen'
at OPERA, and on the other hand that the high energy neutrino looses much of its energy through bremsstahlung 
before it reaches the
OPERA detector. In these works, some 'reasonable' assumptions like energy-momentum conservation, or
that charged mesons obey normal dispersion relations (based on limits on Cerenkov radiation in vacuum),
are made. The point is, that in the new theory all these notions might themselves undergo radical
revisions. Even if some sort of momentum is conserved, it may not be of the form used currently. If
some sort of relativity principle still survives, that will automatically take care of the issue of Cerenkov 
radiation in vacuum. In fact such a relativity principle would forbid the bremsstrahlung process
considered by Cohen and Glashow \cite{kinematics1}. According to \cite{dsr1,dsr2,dsr3} there are indeed
Lorentz violating theories that nevertheless are compatible with a relativity principle. 

In this context, it is worth remembering that the OPERA experiment is really probing \emph{space-time
structure}. Though they talk of neutrino energies and such other notions, these are not central to
an analysis of their data.

Even concepts in which we have great faith like timing need to be reassessed. This is essentially a
so called \emph{Time of Flight}(TOF) method (is it time to rename it as \emph{Time of Fright} method?),
though it is not the traditional 'start time - stop time' TOF method, but is instead a variant with
a \emph{start time distribution} and a \emph{stop time distribution}. TOF data is mostly analysed using classical
methods. But the situation here is highly quantum mechanical, and it is known that \emph{time in quantum
mechanics} can be a very delicate issue (for various perspectives on this, see \cite{qmtime}). There are
various characterstic times like {\emph arrival time, passage time} etc whose meaning in the quantum context
can be ambiguous. In the present circumstance, neutrinos being very weakly interacting, \emph{detection time}
treated quantum mechanically adds further complications. In the end it may well turn out that these nuances
are not particularly relevant for interpreting OPERA data, but it is imperative to get a sound understanding
of these issues given the earth shattering nature of the results.

In a most interesting paper, perhaps the most credible interpretation of the OPERA events, Naumov and Naumov
\cite{wavepacket} have argued that neutrino wavepackets become effectively \emph{oblate spheroids} due to
highly relativistic effects, and that can lead to a \emph{systematic} early arrival times for neutrinos
that are not exactly aligned, at the source, with the detector.

Other subtleties to worry about, regarding the timing of neutrinos at OPERA, are conceptual issues tied
up with the correct way of synchronizing clocks \cite{synchro1,synchro2}. As pointed out by Contaldi \cite{synchro1}, OPERA
uses the so called \emph{one way synchronisation} of clocks, which can be particularly problematic when
synchronizing clocks in non-inertial frames such as the earth (which is non-inertial both because it
is a rotating frame \footnote{I thank Kamal Lodaya for raising the pertinent issue of the experiments
taking place in a rotating frame}, and also because of the earth's gravitational field). According to this analysis,
the lack of synchronization due to time dilation has the correct sign to explain the OPERA results. But building
the lack of synchronization to 60 ns from time dilation, in the manner described, seems unrealistic, but
a careful appraisal is certainly called for.

Before turning to our article, two other proposals, both having to do with the beam structure, that could
seriously impact on the interpretation of the OPERA data, are worth discussing. The first one, due to Henri \cite{beam1},
claims that a fluctuation in neutrino energies of the order of $10\%$ during the roughly 500 ns. of the leading and
trailing edges of the PDF can mimic the alleged superluminal effects claimed by OPERA. While the number of protons
increase and decrease during these phases, the proton energy itself remains at 400 GeV. Thus it is hard to see
the source of this variation in the neutrino energy. Besides, this has to be systematic for the nearly million
extractions, which is hard to accept.

The other, potentially more damaging to the OPERA interpretation, is the claim by Knobloch \cite{beam2} that the 
\emph{Beam Current Tracker}(BCT)
at the CERN end has not registered the true PDF, and that even after several statistical averaging, 30 and 60 ns.
structures are still seen in the PDF as measured by the beam current tracker (see also \cite{brunetti}). If true,
this will cast doubts about the systematics as claimed by OPERA, and could well nullify their claims. It is beyond
our expertise to assess this criticism adequately, and we eagerly await clarification by OPERA.

The plan of this article is as follows: in the next section we recapitulate the most relevant part of the
data on the supernova 1987a. Here we have only looked at the data recorded by the Kamioka detector
\cite{kamioka}. We then analyse a power law dispersion relation for neutrinos, fixing the parameters by requiring
compatibility with the observed \emph{ differential} superluminal neutrino velocity of $2.48\cdot\,10^-5$ in units 
of $c$ at an
average energy of $17\,\, GeV$ \emph{and} the Kamioka data. We then \emph{compute} the expected differentials at
the other energies $13.9\,\, GeV$, $28.1\,\, GeV$ and $42.9\,\, GeV$ where OPERA has data. We compare the calculated values
with the measured values though the statistics of these measurements are not as good as at $17\,\, GeV$. 
We also repeat this exercise by first demanding compatibility between OPERA data at $28.1\,\, GeV$, and
then comparing the calculated values with the data at $13.9\,\, GeV$ and $42.9\,\, GeV$.

In the section after this, we analyse both a flat, energy-independent velocity, as well as a \emph{linear} energy dependence taking into account a possible
\emph{constant shift} term. 
In this section we also analyse a linear energy dependence with a different \emph{shift} term which is taken
to be valid only near OPERA energies. In the subsequent section we discuss the difficulties of a treatment 
that is good both at the OPERA energy scales as well as at the SN1987a energy scales. 
We propose, as an illustrative example, a form
that works at both ends. In the last section we make some concluding remarks.
\section{Essential 1987a data}
\label{1987a}
There is a truly enormous amount of literature discussing various aspects of the supernova explosion
SN1987a in the large magellanic cloud in February 1987. Fortunately for this analysis we need to concentrate
only on a few essentials of the detection by Kamioka \cite{kamioka}. We briefly summarize them in the next para.

The distance to the event, $L_{SN}$, is 1,68,000 light years. A neutrino cluster of 12 events was observed 2 hours
\emph{before} the optical sighting \cite{sighting} of the supernova. The neutrino energies spread over a range $7.5 - 20.0$ 
MeV. In this analysis we have excluded the rather \emph{low} energy event of $6.3\pm1.7$ MeV event arriving
$0.686$ s. after the burst neutrino, and three rather \emph{high} energy events with energies $35.1\pm 8.0,21.0\pm 4.2,
19.8\pm 3.2$ MeV, arriving respectively at $1.541, 1.728, 1.915$ s. after the 
burst event. The exclusion of these \emph{anomalous} events \cite{sn87us} does not affect the reliability of
the analysis given here. The first event was a $\nu_e$ event with $20.0 \pm 2.9$ MeV, identified as a \emph{burst} neutrino.
The subsequent neutrinos, the so called \emph{thermal} neutrinos comprise of neutrinos and antineutrinos
of all flavours. The arrival times of the cluster was spread around 12 seconds, and this accurately established
data is crucial to this analysis. Leaving out the anomalous events, the highest neutrino energy is
around 20 MeV, and the lowest around 7.5 MeV. The average neutrino energy is around 15 MeV.

One should not interpret the early arrival of neutrinos by two hours compared to photons as evidence for any
superluminal neutrino velocity as the photons take longer to escape the opaque stellar medium. This is similar
to light taking a million years longer to escape the solar interrior.

The SN1987a neutrinos are of \emph{electron} flavour, while the OPERA neutrinos are predominantly ${\nu}_\mu$. 
Thus one could take the attitude that there is no common ground, and hence no real conflict between the two.
But oscillations make the situation more complicated, and a reasonable attitude to take is that on the average every
flavour of neutrino would have velocities behaving similarly in both set ups. There are also indications that 
the OPERA effect has to be by and large \emph{flavour independent} if serious troubles with oscillation phenomena
are to be avoided \cite{operasn8}. This provides a justification for the analyses to be discussed next.
\subsection{Analysis based on photon arrival time}
\label{opticalsighting}
As mentioned before the first optical sighting of the supernova was 2 hrs after the neutrino burst \cite{sighting}. It is possible to
use this information to already draw various conclusions of interest. 

The analysis proceeds as follows: let 
$t_{em}^\gamma, t_{em}^\nu$ be the emission times, at the supernova, of photons and neutrinos, respectively. Let us
for the moment assume that photons travel with velocity $c$. The fact that they actually travel slower because of
interstellar dispersion, and due to a possibly very small mass, will only strengthen the conclusions drawn here, not
weaken them. Let $t_{arr}^\gamma, t_{arr}^\nu$ be the respective arrival times. 

If the neutrinos also travel luminally
or subluminally, no interesting conclusions can be drawn. However, if the SN1987a neutrinos also travelled superluminally,
and this can not be ruled out, one can use the optical sighting data to constrain the superluminality. This will be
a \emph{cruder} bound compared to the one to be given in the next subsection, but still useful.

With these assumptions, and an optical sighting $7200$ seconds later, one gets
\begin{equation}
\label{opticalestimate}
t_{em}^\gamma\,+\frac{L_{SN}}{c} = 7200\,+t_{em}^\nu+\,\frac{L_{SN}}{c} -{\delta t}_{SN}
\end{equation}
which can be rearranged as
\begin{equation}
\label{opticalestimate2}
{\delta t}_{SN} = 7200\,+t_{em}^\nu -t_{em}^\gamma\,
\end{equation}
Since under no circumstances can the photons leave the supernova \emph{earlier} than the neutrinos, one gets
\begin{equation}
\label{opticalestimate3}
{\delta t}_{SN} \le 7200
\end{equation}
seconds. This is of course being ultra-conservative. One may realistically take photons to have left one hour
later and get
\begin{equation}
\label{opticalestimate3}
{\delta t}_{SN} \le 3600\quad\quad \frac{{\delta v}_{SN}}{c}\,\le \frac{3600\,s.}{168000\,y.} = 6.8\cdot\,10^{-10}
\end{equation}
\subsection{Analysis based on neutrino arrival times}
\label{neutrinoarrival}
The analysis now proceeds as follows: let 
$t_{em}^{high}, t_{em}^{low}$ be the emission times, at the supernova, of neutrinos of the highest and lowest
energies, respectively. 
Let $t_{arr}^{high}, t_{arr}^{low}$ be the respective arrival times. One now gets
\begin{eqnarray}
\label{neutrinoestimate}
t_{arr}^{low} &=& t_{em}^{low}+\,\frac{L_{SN}}{c}-\,\frac{L_{SN}{\delta v}_{SN}^{low}}{c^2}\nonumber\\ 
t_{arr}^{high} &=& t_{em}^{high}+\,\frac{L_{SN}}{c}-\,\frac{L_{SN}{\delta v}_{SN}^{high}}{c^2} 
\end{eqnarray}
leading to
\begin{equation}
\label{neutrinoestimate2}
\frac{L_{SN}}{c^2}\{{\delta v}_{SN}^{high}\,-{\delta v}_{SN}^{low}\}\le\,[12+(t_{em}^{high}-t_{em}^{low})]
\end{equation}
Unlike the photon arrival case, no simple arguments exist for $(t_{em}^{high}-t_{em}^{low})$, and it would
require details of neutrino emission mechanisms from supernovae(to get some idea of these issues see \cite{lamb,raffelt,nuemission1,nuemission2,nuemission3}).
In most scenarios considered in these works, high energy neutrinos precede low energy ones. It seems very reasonable to take the quantity $[..]$ in eqn.(\ref{neutrinoestimate2}) to be about 10 seconds. This
leads to
\begin{equation}
\label{neutrinoemission3}
\{{\delta v}_{SN}^{high}\,-{\delta v}_{SN}^{low}\}\le\,2.0\cdot\,10^{-12}\,c
\end{equation}
It is also important to consider the IMB data \cite{imb} in a more careful analysis. The energies of the detected
neutrinos are higher in IMB. If very accurate timing was available for all detectors, that also could have been
used to constrain the energy dependence of ${\delta v}_{SN}$. But unfortunately, the Kamioka timing had an
uncertainty of $\pm 1$ min; the IMB had by far the most accurate timing record, with an error of $\pm 50$ milliseconds.
A determination of the so called \emph{offset} time from fitting data to models of neutrino emission seems
to indicate that the first IMB neutrino arrived $0.1$ s. before the first Kamioka neutrino, in the so called
two component model of neutrino emission \cite{nuemission3}. But given the complexity of the issues, too much
significance should not be attached to this. An intriguing suggestion has been made by Fargion \cite{fargion}
to search past records of IMB for evidence supporting superluminal behaviour of neutrinos.

If a more detailed analysis of arrival times could be made on the basis of a soundly established model of neutrino
emission from the supernova, one could put even more stringent limits than the one in eqn.(\ref{neutrinoemission3}),
but the situation here is very complex (see for example \cite{lamb,raffelt,nuemission1,nuemission2,nuemission3}).

It should be emphasized that both eqn.(\ref{opticalestimate3}) and eqn.(\ref{neutrinoemission3}) only reflect the
possible upper limits to ${\delta v}_{SN}$. They do not imply that supernova neutrinos are superluminal, but that
the available data are indeed \emph{consistent} with them being so. They may in the end turn out to be subluminal, or
just luminal too.
We have left out any discussion of the effect of neutrino masses. The current thinking on these masses is that
they can at most be of a few eV with one of them even being massless. For 20 MeV neutrinos, the correction to
$\frac{\delta v}{c}$ is $\simeq\,\frac{1}{2}\cdot\,10^{-14}$, if the mass is taken to be $2 eV$.
\section{General power law}
\label{powerlaw}
Now we consider a possible power law dependence for $\delta v$. This analysis should be taken as an
exercise in phenomenology with no pretenses to any deeper theoretical considerations. For integer powers, these
belong to the class of so called \emph{Distorted Special Relativistic} theories introduced in \cite{dsr1,dsr2,dsr3}. OPERA data
has been analysed in the context of these models in \cite{camelia}. An interesting point stressed in \cite{camelia}
is that these DSR type theories still allow for a relativity principle.

The simplest
candidate for a power law behaviour is
\begin{equation}
\label{powerlaw}
\frac{{\delta v}}{c} = \alpha_1\,E^m
\end{equation}
where $\alpha_1$ is a constant with dimensions of $E^{-m}$. We shall require eqn.(\ref{powerlaw}) to be valid at all
energies. Therefore
\begin{equation}
\label{powerlaw2}
\frac{{\delta v}_{SN}}{{\delta v}_{OP}} = (\frac{E_{SN}}{E_{OP}})^m
\end{equation}
The data from OPERA for the advancement times are: $60.7,\,60.3,\,53.1,\,67.1\,$ nanoseconds at $17.0,\,28.1,\,13.9,\,42.9\,$ 
GeV respectively. In terms of $\frac{{\delta v}}{c}$ these correspond to $(2.48,\,2.46,\,2.17,\,2.74)\cdot\,10^{-5}$
respectively. The statistical errors are the least at $17$ GeV, getting progressively higher at the other energies.
The quoted statistical errors are : $(\pm\,0.28,\,\pm\,0.54,\,\pm\,0.77,\,\pm\,0.75)\cdot\,10^{-5}$ respectively, and the systematic
errors are $\pm\,0.30\cdot\,10^{-5}$ \cite{actone}.
\subsection{Estimating $\alpha_1,\,m$ from photon arrival time data}
We shall estimate the parameters $\alpha_1,m$ first using the photon arrival time data. 
Taking average energy of SN1987a at 15 Mev and the OPERA energy at 17 GeV, where the statistics are the best, the energy
ratio comes out to be $8.8\cdot\,10^{-4}$. Taking the estimate of eqn.(\ref{opticalestimate3}) for ${\delta v}_{SN}$
one gets, on using eqn.(\ref{powerlaw2}),
\begin{equation}
\label{powerlaw3}
m = 1.5 \quad\quad \alpha_1 = (20\, TeV)^{-1.5}
\end{equation}
Using eqn.(\ref{powerlaw3}), one can estimate ${\delta \beta} = \frac{{\delta v}}{c}$, in units of $10^{-5}$, at the energies
of $13.9,\,28.1,\,42.9$ GeV :
\begin{equation}
\label{powerlaw4}
{\delta \beta}(13.9) = 1.8 \quad\quad {\delta \beta}(28.1) = 5.2 \quad\quad {\delta \beta}(42.9) = 10.0
\end{equation}
respectively. 
Thus the estimates based on the powerlaw fit of eqn.(\ref{powerlaw}) are off
the experimentally measured central values by $.5,\,9,\,10$ standard deviations(statistical) respectively. Thus this power
law fit works rather poorly.

This can be understood from the fact that the values at $17 GeV$ and $28.1 GeV$ are very close. In fact, they are
consistent with no energy dependence at all. It is therefore instructive to make an analysis with the data at $13.9,\,28.1,\,42.9$
GeV only. Of course the statistical errors are much higher, and this may look like deliberate data selection, but its
worth going through it given the scantiness of data at different energies.
An analysis identical to the one above, now gives
\begin{equation}
\label{powerlaw5}
m = 1.39 \quad\quad \alpha_1 = (50\, TeV)^{-1.5}
\end{equation}

Using eqn.(\ref{powerlaw5}), one can estimate ${\delta \beta} = \frac{{\delta v}}{c}$, in units of $10^{-5}$, at the energies
of $13.9,\,42.9$ GeV :
\begin{equation}
\label{powerlaw6}
{\delta \beta}(13.9) = .93 \quad\quad {\delta \beta}(42.9) = 4.46
\end{equation}
respectively. Now the estimates based on the powerlaw fit of eqn.(\ref{powerlaw}) are off
the experimentally measured central values by $1.63,\,2.3$ standard deviations(statistical) respectively. As a last bit
of comparison, we estimate ${\delta \beta}$ at $17 GeV$ that eqn.(\ref{powerlaw5}) would give: ${\delta \beta}(17) = 1.39\cdot\,10^{-5}$,
differing by the measured value by $3.9$ standard deviations.

\subsection{Estimating $\alpha_1,\,m$ from neutrino arrival time data}
In a similar manner we can estimate the parameters $\alpha_1,\,m$ by using the estimates based on neutrino
arrival times as given in eqn.(\ref{neutrinoemission3}). As input, we need ${\delta v}_{SN}$ at the highest
energy, 20 MeV. Getting
it from eqn.(\ref{neutrinoemission3}) requires an estimate for ${\delta v}_{SN}$ at the lowest energy,
taken here to be 7.5 MeV, which can be obtained
as a fraction of the former if we already know the exponent $m$. We shall circumvent this in a practical manner
by simply taking the rhs of eqn.(\ref{neutrinoemission3}) to be the estimate for ${\delta \beta}_{SN}$ at 20 MeV.

As before, using this estimate for ${\delta \beta}_{SN}$ in eqn.(\ref{powerlaw2}) gives
\begin{equation}
\label{powerlaw7}
m = 2.42 \quad\quad \alpha_1 = (1.4\, TeV)^{-2.42}
\end{equation}
Using eqn.(\ref{powerlaw7}), one can estimate ${\delta \beta} = \frac{{\delta v}}{c}$, in units of $10^{-5}$, at the energies
of $13.9,\,28.1,\,42.9$ GeV :
\begin{equation}
\label{powerlaw8}
{\delta \beta}(13.9) = 1.52 \quad\quad {\delta \beta}(28.1) = 8.4 \quad\quad {\delta \beta}(42.9) = 23.3
\end{equation}
respectively. 

Thus the neutrino arrival times give a much steeper power law than the one given by photon arrival times. It is
therefore not surprising that the estimated results are much worse compared to those given by
eqn.(\ref{powerlaw4}). As before, we give the results that one would get if one had only used the data 
at $13.9,\,28.1,\,42.9$ GeV:
\begin{equation}
\label{powerlaw9}
m = 2.11 \quad\quad \alpha_1 = (4.4\, TeV)^{-2.11}
\end{equation}
Using eqn.(\ref{powerlaw9}), one can estimate ${\delta \beta} = \frac{{\delta v}}{c}$, in units of $10^{-5}$, at the energies
of $13.9,\,42.9$ GeV :
\begin{equation}
\label{powerlaw10}
{\delta \beta}(13.9) = .56 \quad\quad {\delta \beta}(42.9) = 6.00
\end{equation}

Quadratic corrections of the type $m=2$ have been suggested countless number of times in the so called \emph{Quantum Gravity
} inspired models of dispersion relations. If OPERA data even at high energies is taken seriously, these are seen
to perform rather poorly. Furthermore, the intrinsic energy scales in these models is the \emph{Planck Mass}, but
in these fits that scale is turning out to be around TeV. But then, there are also models where the 'Planck
Mass' can be much lower.

The lesson one learns from the analysis so far is that while fitting SN1987a requires \emph{steeper} energy
dependence, the OPERA data on energy dependence prefers \emph{flatter} dependence. Striking a balance between
these opposing trends is the key to reconciling the OPERA and SN1987a data.

\section{Flatter Energy Dependences.}
\subsection{Energy Independent Superluminality.}
\label{flat}
If one takes both the data of 60.7 nanoseconds at 17 GeV, as well as the 60.3 nanoseconds at 28.1 GeV, one
has to more or less conclude that the observed superluminality is \emph{energy independent}:
\begin{equation}
\label{flatdelta}
{\delta \beta}_{OP} \simeq\,2.48\cdot\,10^{-5}
\end{equation}
This is what the dispersion relations from the so called \emph{Standard Model Extensions} \cite{kostelecky1,
kostelecky2,diazreview} as well as the Coleman-Glashow analysis \cite{colemanglashow} would give. These proposals
were made in the context of the so called \emph{High Energy Violations} of Lorentz invariance. But for neutrinos,
these analyses would seem to imply their validity as long as neutrino energies are much higher than neutrino masses.
Hence they should be applicable to the SN1987a neutrinos too. 

A flat velocity surplus of eqn.(\ref{flatdelta}) would
imply that the supernova neutrinos preceded the photons by about 4 years! While neutrinos hitting the earth 4 years before
the photons can not be ruled out \cite{operasn8}, the fact that the neutrinos observed by Kamioka almost fully accounted for the
energy output in neutrinos that is expected from the standard supernova models, would cause some problems of energetics.
But it is fair to say that one needs a complete analysis of neutrino oscillations with the superluminality of
eqn.(\ref{flatdelta}) built in before one can make confident statements. There are already such analyses of neutrino oscillations
available in the literature \cite{diazreview,joao} . But this line of thinking requires the
above equation to be valid at all energies. But we shall argue that it is established, if at all, only at OPERA
energies, and we shall propose interpolations that can comfortably sit with the SN1987a data presented earlier.

\subsection{Linear Energy Dependences.}
It is interesting that the central values of ${\delta t}$ quoted by OPERA at 13.9 and 42.9 GeV, namely, 53.1 ns and
67.1 ns, when combined with 60.3 ns at 28.1 GeV, suggest a rather good \emph{linear} fit to the data. But from our
analysis till now, a flatter energy dependence at OPERA energy scales does not extrapolate well to the supernova
scales. In fact the linear dependence suggested by the above mentioned data is roughly 1 ns increase for every 2 GeV.
Extrapolating 53.1 ns at 13.9 GeV to supernova energy scales (nearly zero on OPERA energy scales), one still gets
46 ns which translates to a $\frac{{\delta v}}{c}$ of $\simeq\,2\cdot\,10^{-5}$. This is unacceptably large,
leading to neutrinos arriving nearly 3.4 years before photons(see, however, \cite{operasn8})!

A simple way of circumventing this is to consider a linear dependence with a suitable \emph{offset}
\begin{equation}
\label{offset}
{\delta \beta}_{OP} = \alpha_2\,E\,+\,{\delta \beta}_0
\end{equation}
where ${\delta \beta}_0$ is so adjusted as to cancel most of the large value mentioned before. In eqn.(\ref{offset}), $\alpha_1$
is roughly $4.0\cdot\,10^{-7}$ per 2 GeV. Even without
working out ${\delta \beta}_0$, we can see that this too ends up in trouble. This is because
\begin{equation}
\label{offsetprob}
{\delta \beta}_{SN}^{high}\,-{\delta \beta}_{SN}^{low} \simeq \alpha_1\cdot\,12.5 MeV \simeq 2.5\cdot\,10^{-9}
\end{equation}
is still very large, predicting a differential time of arrival of nearly two hours for these neutrinos,
whereas Kamioka data constrains them to be at most 12 seconds.

Eqn.(\ref{offset}), with just the offset term, is what the Coleman-Glashow like Lorentz violating dispersion relation
\cite{kostelecky1,kostelecky2,colemanglashow} would give. 
Thus unless one interprets the
Coleman-Glashow form to be valid only for energies well above a GeV scale, we see that it does not work for neutrinos.
Eqn.(\ref{offset}) is an example of the so called DSR type theories discussed by \cite{dsr1,dsr2,dsr3}. Among the many 
Lorentz violating theories, these may have the advantage that they may 
still accommodate a relativity principle.

As it appears to be very difficult to reconcile both OPERA and SN1987a data with a \emph{single scale}
energy dependence, let us abandon that approach and first seek an energy dependence valid close to
the OPERA energies (namely 10 to 40 GeV) only. Then, a very nice fit to the available data is provided
by eqn.(\ref{offset}) with $\alpha_2 = (5\cdot\,10^3\,TeV)^{-1}$ and $\frac{{\delta v}_0}{c} = 1.91\cdot\,10^{-5}$,
that is
\begin{equation}
\label{offset2}
{\delta \beta}_{OP} = 2\cdot\,10^{-7}\,\frac{E}{GeV}\,+\,1.91\cdot\,10^{-5}
\end{equation}

A comparison of OPERA data with various forms of dispersion relations has also been done in \cite{camelia}.
Their comparison with SN1987a is rather preliminary, and not as detailed as the one given here. They even
hint at some sort of a \emph{metatheorem}(!) claiming that it would not be possible to achieve consistency
with both OPERA and SN1987a data. We do not agree with that. Many papers have appeared that have analyzed the
compatibility issues between OPERA and SN1987a \cite{operasn1,operasn2,operasn3,operasn4,operasn5,operasn6,operasn7,operasn8,operasn9}.

\section{Interpolating OPERA and SN1987a}
\label{interpolate}
Thus, if we want to give an energy dependence formula, with a single \emph{intrinsic scale}, that fits
both OPERA and SN1987a data, it is almost impossible. If , however, one allows another intrinsic scale,
many possibilities open up. We do not wish to explore all possible such forms, as such an exercise is
somewhat pointless at the present juncture. We give one of the simplest forms:
\begin{equation}
\label{interpolate}
{\delta \beta}\,=\,{\delta \beta}_{OP}(\frac{E_{GeV}^2}{E_{GeV}^2\,+\,M_L^2})^n
\end{equation}
where ${\delta \beta}_{OP}$ is from either eqn.(\ref{flatdelta}) or eqn.(\ref{offset2}), and $M_L$ in GeV is a scale which 
is much larger than supernova energies, but much smaller than OPERA energies.
The parameters $M_L,\,n$ are constrained by SN1987a data according to
\begin{equation}
\label{interpolate2}
{\delta \beta}_{SN}\,\simeq 2.0\cdot\,10^{-5}\,(\frac{E_{GeV}^2}{M_L^2})^n
\end{equation}
We have used $\simeq$ here because the prefactor in eqn.(\ref{interpolate2}) is 2.48 if we had used eqn.(\ref{flatdelta}), and
1.91 if we had used eqn.(\ref{offset2}). This hardly makes much difference to the estimates of $M_L,\,n$. We will see in the next subsections that estimates of eqn.(\ref{opticalestimate3}) and eqn.(\ref{neutrinoemission3})
quite severely constrain the parameters in eqn.(\ref{interpolate2}).
\subsection{Photon arrival constraints}
Let us begin with the estimate as given in eqn.(\ref{opticalestimate3}) and take it to be for 15 MeV neutrinos. It
is straightforward to work out that for $n=1,2$ the scale $M_L$ works out to $2.5,\, 0.2 $ GeV respectively.
Both these scales satisfy our criterion that the scale should be much smaller than OPERA energies, and much larger than
supernova neutrino energies. At this stage there is no other criterion to choose one over another.
\subsection{Neutrino arrival constraints}
If instead of eqn.(\ref{opticalestimate3}) we use eqn.(\ref{neutrinoemission3}), we get that for $n=1,2$ the
scale $M_L$ is $50, 1$ GeV respectively. Clearly, $M_L=50$ GeV is unacceptable, while $M_L=1$ GeV is certainly
acceptable. In any case, one should be using the more accurate and cleaner estimates provided by 
eqn.(\ref{neutrinoemission3}), and rule out $n=1$ in eqn.(\ref{interpolate2}).
\section{Discussion and conclusions}
\label{conclusions}
Our aim was to discuss possible energy dependence of the superluminal neutrino velocities reported by the
OPERA collaboration. Our motivations are three fold: i. an understanding of energy dependence will provide
an important key to unraveling the mysteries, ii. without any energy dependence it is \emph{impossible}
to reconcile the findings of OPERA with the accurate timing data from SN1987a, and finally, iii. if OPERA data
at both 17 and 28.1 GeV are taken seriously, then, near OPERA energies the velocities are energy-independent.
But the mean values they quote for 13.9, 28.1 and 42.9 GeV hint at a clear energy dependence.
Though the statistical errors for the measurements at 13.9 and 42.9 GeV are much poorer, as the sample sizes were
drastically reduced in number due to the reliance on only CC events, we have taken the central values, which 
are significant at three standard deviations, to parametrize the energy dependence on the one hand, and to
reconcile them with SN1987a on the other.

For the SN1987a data, we have used both the timing of the first optical sighting, as well as the time spread
in the cluster of 12 events observed at the Kamioka detector. We have first attempted a power law fit
to both the OPERA data at 17 GeV as well as the SN1987a estimates. Fitting the light arrival time gives an
intrinsic mass scale of $20\, TeV$ and an energy exponent of 1.50. The calculated values at 13.9, 28.1 and 42.9 GeV
from this fit yield values that deviate from the quoted mean values by .5, 5.2 and 10 standard deviations(statistical).
On the other hand, fitting photon arrival times to OPERA data at 28.1 GeV yields a scale of 50 TeV, and an
exponent of 1.39.

A fit based on the 12 seconds clustering of the neutrino events at Kamioka iand 17 GeV data from OPERA gives, on the other hand, an
intrinsic mass scale of $1.4\,TeV$ and an exponent of $2.42$, which could be interpreted as really
$2.0$ given the tentativeness of the estimates. Extrapolating these to 13.9, 28.1 and 42.9 GeV, one finds
the calculated values of ${\delta \beta}$ to be 1.52, 8.4 and 23.3. Such
dependences with $m=2$ are suggested by many \emph{quantum gravity inspired} models, but the scales found here are very
different.

Motivated by the near linear increase of ${\delta t}$ reported by OPERA for the observations at 13.9, 28.1
and 42.9, we next attempted a linear fit but with an offset so that the linear behaviour extrapolated to
supernova energies would reconcile with the supernova arrival limits. If the slope is determined by the OPERA data,
it is found that irrespective of the offset, one gets a differential arrival time for 7.5 MeV and 20.0 MeV
supernova neutrinos that is very clearly in conflict with the arrival times.

Interpreting all this as the inability of any single-scale energy dependence to account for both OPERA
and SN1987a data, we decided to check how well a linear fit will work only when restricted to OPERA energies.
A mass scale of $5\cdot\,10^{3}$ TeV and an intercept(offset) of $2.0\cdot\,10^{-5}$ was obtained. This fit
works very well. This can be 
tested by measuring ${\delta t}$ at other, but comparable, energies as OPERA.

Finally, an interpolation was sought between this linear behaviour valid near OPERA energies, and the upper
limits on neutrino velocities obtained from supernova data. The important point to notice is that these
supernova arrival time limits are still consistent with superluminal velocities for supernova neutrinos.
This is an exciting possibility that needs to probed further. There are clearly many such interpolations
possible. One of them is given in eqn.(\ref{interpolate2}). The additional scale $M_L$ and the
exponent $n$ are again
constrained by arrival time data and the following,
\begin{equation}
\label{interpolate3}
{\delta \beta}\,=\,{\delta \beta}_{OP}(\frac{E_{GeV}^2}{E_{GeV}^2\,+\,1.0})^2
\end{equation}
seems most probable. Thus, this analysis anticipates an $E^4$- dependence for ${\delta v}$ near supernova
energies. Even at $1\, GeV$, this relation predicts a $\frac{\delta v}{c}$ which is only $\frac{1}{5}$th
of that reported by OPERA at 28.1 GeV! 
\subsection{Further Work}
The OPERA collaboration should increase the statistics of the measurements done with the CC events. Determining
reliably the energy dependence of the effect, if true, is one of the most important things to be done. Efforts
should also be made to measure the effect at different energies. An accurate determination of the energy distribution
of the neutrino flux is also important.

It is important also to perform this experiment at different baseline lengths because it is crucial to separate
effects of production and detection from those of propagation(the effects discussed by \cite{beam1} are an example of
the kind that have no dependence on the baseline distance). The claim of 18cm accuracy in the baseline length
is central to the establishment of the superluminality. It is therefore critical to find several independent
checks on what could easily become the \emph{Achilles heel} for the experiment.

As already emphasized by Coleman and Glashow \cite{colemanglashow}, the superluminal velocity effects and
the energy dependence of them can have novel manifestations in oscillation phenomena. It is therefore of
utmost importance to remodel oscillation calculations by including the energy dependences considered here,
and then subject them to experimental tests. In principle, the various parameters that entered the
interpolation formula in eqn.(\ref{interpolate2}) could be \emph{flavour} dependent. Even the TOF analysis
itself should be remodelled taking into account possible flavour dependences, as well as effects of oscillations.

The supernova neutrino data, from all the detectors that observed them, also needs to be thoroughly reexamined,
particularly from the points of view of energetics and timing.

The most challenging task, should OPERA results be found to stand further scrutiny, will be in arriving at a reappraisal
of our notions of space and time. The entire edifice of theoretical physics stands on these pillars.

\subsection{Acknowledgements}
{The author thanks Bala Sathiapalan, Kalyana Rama, M.V.N. Murthy, R. Shankar, H.S. Mani and Kamal Lodaya for many useful discussions.
He is thankful to Kalyana Rama for stressing the importance of energy dependences.
He acknowledges support from
Department of Science and Technology to the project IR/S2/PU-001/2008.  }

\end{document}